\documentstyle[11pt,newpasp,twoside,epsf]{article}
\def\edcomment#1{\iffalse\marginpar{\raggedright\sl#1\/}\else\relax\fi}
\reversemarginpar

\begin{document}
\title{Reverberation Measurements of Quasars and the
Size-Mass-Luminosity Relations in Active Galactic Nuclei}
\author{Shai Kaspi}
\affil{Department of Astronomy and Astrophysics, The Pennsylvania State
University, University Park, PA 16802, USA}

\begin{abstract}
A 7.5 years spectrophotometric monitoring program of 28 Palomar-Green
quasars to determine the size of their broad emission line region (BLR)
is reviewed. We find both the continuum and the emission line fluxes of
all quasars to vary during this period. Seventeen objects has adequate
sampling for reverberation mapping and in all of them we find the
Balmer line variations to lag those of the continuum by $\sim$100
days.  This study increases the available luminosity range for studying
the size--mass--luminosity relations in active galactic nuclei (AGNs)
by two orders of magnitude and doubles the number of objects suitable
for such studies. Combining our results with data available for Seyfert
1 galaxies, we find the BLR size to scale with the rest-frame
5100\,\AA\ luminosity as $L^{0.70\pm 0.03}$. This result is different
from previous studies, and suggests that the effective ionization
parameter in AGNs may be a decreasing function of luminosity. We are
also able to constrain, subject to the assumption that gravity
dominates the motions of the BLR gas, the scaling relation between the
mass of the central black hole and the AGN's luminosity. We find that
the central mass scales with the 5100~\AA\ luminosity as $M\propto
L^{0.5 \pm 0.1}$.

A program to monitor 11 high-luminosity quasars is presented here for
the first time. Preliminary results from this program indicate
continuum variation of order of 0.1 mag in all objects. We illustrate
the importance and feasibility of monitoring those objects
spectrophotometrically. When this program will be completed
reverberation mapping studies will cover the entire AGNs luminosity
range from $10^{41}$ to $10^{48}$\,ergs\,s$^{-1}$.
\end{abstract}

\section{Introduction}
\label{introduction}

Broad emission lines in active galactic nuclei (AGNs) emerge from the
innermost regions of these objects. As such, they provide important
information (e.g., composition, dynamics, physical conditions, and
geometry) about the AGNs' unresolved regions. Reverberation mapping,
observing the degree and nature of the correlation between continuum
and emission-line flux variations, is one of the major tools for
studying the distribution and kinematics of the gas in the broad line
region (BLR) and to study the central masses of AGNs (e.g., Peterson
1993; Netzer \& Peterson 1997). At a first approximation reverberation
mapping yields a measure for the size of the BLR. Combining this size
with the line profile, which represent the kinematics, one can estimate
the mass of the central source in an AGN. Determining this mass is most
important for the understanding and modeling of the AGNs phenomenon.
Establishing the relations between the physical properties (such as BLR
size, mass of the central source, and the AGN luminosity) of an AGNs
sample will provide us with important information for understanding the
characteristics which are common to all AGNs.

In the past $\sim 15$ years about 17 low-luminosity AGNs (Seyfert~1
galaxies) have been successfully monitored and produce statistically
meaningful BLR sizes (see Wandel, Peterson, \& Malkan 1999, and
references therein). Best studied among these is the Seyfert 1 galaxy,
NGC~5548, which was monitored from the ground for over eight years, and
from space for several long periods (Peterson et al. 1999, and
references therein). Several other Seyfert~1s were observed for periods
of order 1 year or less, and nine Seyfert 1s were studied over a period
of eight years (Peterson et al. 1998a). The measured time lags between
the emission lines and the continuum light curves in these objects can
be interpreted in terms of the delayed response of a spatially-extended
BLR to a variable, compact source of ionizing radiation. While the
observations do not uniquely determine the geometry of the BLR, they
give its typical size which, for Seyfert~1 galaxies, is of the order of
light-days to several light-weeks ($\sim 10^{16}$--$10^{17}$\,cm).
Recent studies have shown that the time lags determined in NGC 5548 for
different observing seasons correlate with the seasonal luminosity of
the object (Peterson et al. 1999), and have presented evidence for
Keplerian motions of the BLR gas (Peterson \& Wandel 1999).

While there has been great progress in mapping Seyfert 1's few similar
studies of the more luminous AGNs -- the quasars -- have been
presented. There still have been some open questions such as, do quasar
emission lines respond to the continuum changes, as seen in Seyfert
galaxies? What is the relative amplitude of the response? What is the
lag of the response, reflecting the light-travel time across the BLR?
Do quasar BLR sizes scale with AGNs luminosity, and lie on a continuous
relation from the faintest Seyferts to the bright quasars?

There are several difficulties when attempting to monitor
high-luminosity AGNs. Quasars have fainter apparent magnitudes, hence
one needs larger telescopes and/or much longer integration limes. Since
quasars variability time scales are longer than Seyferts 1's and their
BLRs are expected to be an order of magnitude larger than in Seyfert
1's, we need much longer monitoring periods. When monitoring Seyfert
1's one often use the narrow emission lines to intercalibrate the
observed spectra, however, in quasars narrow emission lines are very
faint (or not present at all) and a different relative flux calibration
method needs to be exploit.

Past attempts to spectrophotometrically monitor quasars have generally
suffered from temporal sampling and/or flux calibrations that are not
sufficient for the determination of the BLR size (e.g., Zheng 1988;
P\'{e}rez, Penston \& Moles 1989; Korista 1991; O'Brien \& Gondhalekar
1991; Jackson et al. 1992; Koratkar et al. 1998; Wisotzki et~al.
1998).  The quasar best studied by IUE, 3C\,273, has yielded disputed
results when different researchers have analyzed similar IUE monitoring
data sets. Both O'Brien \& Harries (1991) and Koratkar \& Gaskell
(1991a) found a measurable and similar lag between continuum and BLR
variations, while Ulrich, Courvoisier, \& Wamsteker (1993) argued that
the line variations reported in the earlier studies were only
marginally significant.

As a more definite results on the BLR size in quasars is needed we
began two quasars' reverberation mapping programs on which we report in
this contribution. The first program is the monitoring of a sub-sample
of 28 Palomar-Green (PG) quasars. The program and its results are
described in details by Kaspi et al. (2000) and references therein.
Here we present the basic concepts and results of this program and
discuss the size--mass--luminosity relations in AGNs.  The second
program of monitoring 11 high-luminosity high-redshift quasars is
reported here for the first time and some preliminary results are
presented.

\section{The PG quasars -- Sample \& Observations}
\label{observations}

In contrast to Seyfert 1 galaxies, which were chosen for reverberation
mapping mainly because of their variability properties, the quasars in
our sample were selected according to their observable properties. The
sample of 28 quasars is drawn from the 114 PG quasars (Schmidt \& green
1983). The PG quasars are well studied objects over the whole
electromagnetic spectrum from X-ray to Radio (e.g., Neugebauer et al.
1987; Kellermann et al. 1989; Boroson \& Green 1992; Brandt, Laor, \&
Wills 2000). Many of their observational properties are well known and
adding the information about their variability properties and BLR sizes
will increase our understanding of the AGNs phenomenon.

We selected objects with northern declination, $B<16$ mag, redshift
$z<0.4$ (so that the Balmer lines can be observed in the optical
region), and a bright comparison star within 3{\farcm}5 of the quasar.
The absolute magnitude range covered by the sample is $-23>M_B>-27$ mag
(using $q_0=0.5$, $H_0=75$ km\,s$^{-1}$\,Mpc$^{-1}$, and zero
cosmological constant) and the bolometric luminosity range is
$4\times10^{44}<L<3\times10^{46}$ ergs\,s$^{-1}$.

Spectrophotometric observations of the sample were carried out form
1991 March, over a period of 7.5 years, until September 1999. We used
the Wise Observatory (WO) 1\,m telescope to observe the sample once a
month (when the objects are observable) and the Steward Observatory
(SO) 2.3\,m telescope once every $\sim 4$ months. The spectral range is
from 4000 to 8000 \AA\ with spectral resolution of $\sim 10$ \AA. The
Spectrophotometric calibration for each quasar was accomplished by
rotating the spectrograph slit to the appropriate position angle so
that the nearby bright star was observed simultaneously with the
quasar. A wide slit was used to minimize the effects of atmospheric
dispersion. This technique provides excellent calibration even during
poor weather conditions, and accuracies of order 1\%--2\% can easily be
archived.

Alongside the spectrophotometric monitoring we carried out at the WO
1\,m telescope a broad band photometric monitoring in $B$ and $R$. The
28 quasars in our sample were included in a sample of 45 PG quasars
which were monitored monthly to find the variability properties of the
PG quasars (Giveon et al. 1999). We used the photometric data (by means
of differential photometry with other stars in each field) to check on
the continuum behavior found in the the spectrophotometric monitoring
and to verify that non of our comparison stars are variables to within
$\sim$2\%. We also used the photometric observations to add additional
epochs to the continuum light curves of the quasars.

\begin{figure}
\vglue0.5cm
\hglue1.0cm{\epsfxsize=11.2cm\epsfbox{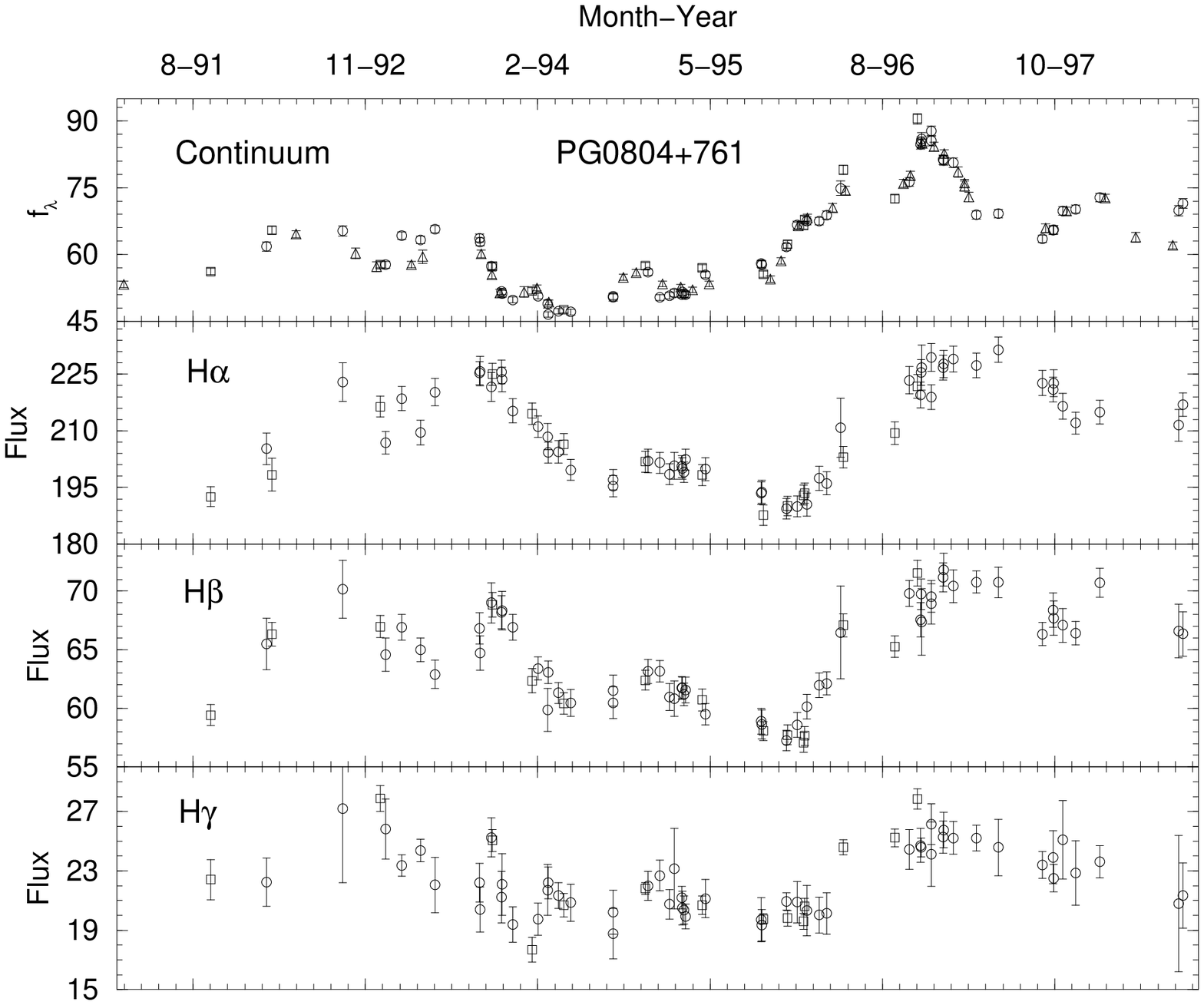}}
\vglue-0.05cm
\hglue1.2cm{\epsfxsize=11cm\epsfbox{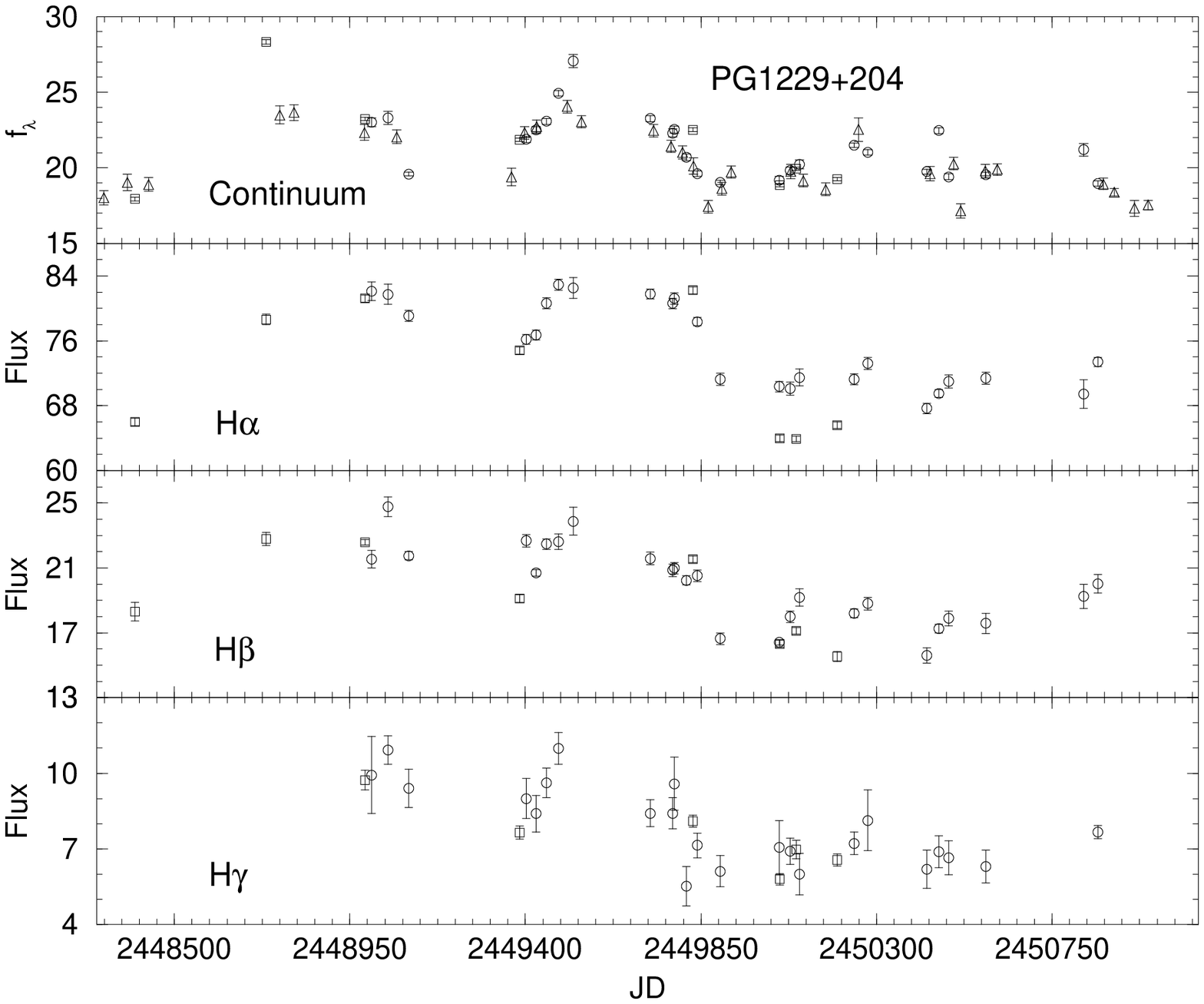}}
\vglue-0.685cm
\caption{Light curves for two PG quasars. Circles are
spectrophotometric data from WO, squares are spectrophotometric data
from SO, triangles are photometric data from WO. Continuum flux
densities, f$_{\lambda}$, are given in units of 10$^{-16}$\,{ergs\,s$^{-1}$\,cm$^{-2}$\,\AA$^{-1}$}.
Emission-line fluxes are displayed in units of 10$^{-14}$\,{ergs\,s$^{-1}$\,cm$^{-2}$}.
Horizontal axis given in Julian Day (bottom) and UT date (top).}
\label{light_curves}
\end{figure}

Continuum light curves were extracted at rest wavelength of about 5100
\AA\ and line light curves of H$\alpha$, H$\beta$, and H$\gamma$ (where
available) for all objects. The photometric data were combined into
the continuum light curves. Light curves for two quasars are shown in
Figure~\ref{light_curves}. PG\,0804+761 is our best sampled object with
$\sim70$ spectroscopic observations and $\sim 40$ photometric
observations. The light curves of this object clearly show the
variations of the three Balmer lines to lag the continuum variations.
Also they demonstrate how small variations in the continuum light curve
smear out in the line light curves -- this effect is a result of a
stratified BLR. PG\,1229+204 is a more typical example of our sample.
It has $\sim 30$ spectroscopic observations and $\sim 30$ photometric
observations. In this object it is hard to identify the time lag by
looking at the light curves but the use of cross-correlation is clearly
yielding a time lag (see next section).

\section{Observational Results and Time Lag Determination}

At the end of the 7.5 years monitoring period we find that for 17
objects out of the 28 there are more than 20 spectroscopic
observations. Typically there are between 20 to 70 observations for
each of these 17 quasars. The other 11 quasars have less than 10
spectroscopic observations (the typical number is 5 observations) which
is not adequate sampling for time series analysis. These 11 objects are
excluded from further discussion hereafter.

All the 17 quasars with adequate spectroscopic sampling had gone
continuum variation which, quantified as $F_{\rm max}/F_{\rm min}-1$, lie in
the range of 35\%--150\% for different objects. All 17 quasars also show
line flux variation which follow the continuum variation with an
amplitude of about half of the continuum variations (see
Figure~\ref{light_curves}).

\begin{figure}
\vglue-0.3cm
\hglue0.8cm{\epsfxsize=11.5cm\epsfbox{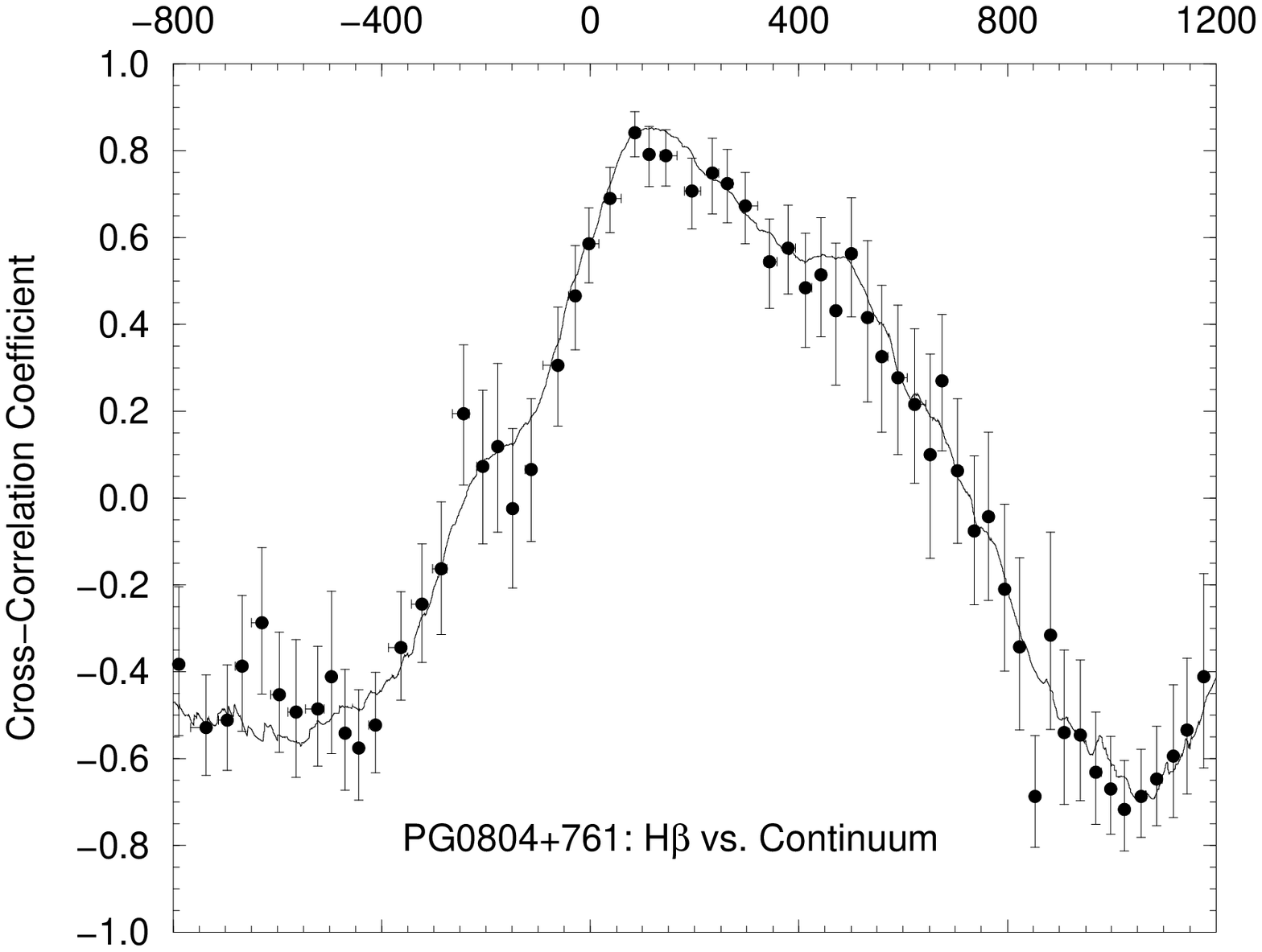}}
\vspace{-0.3cm}
\hglue0.8cm{\epsfxsize=11.5cm\epsfbox{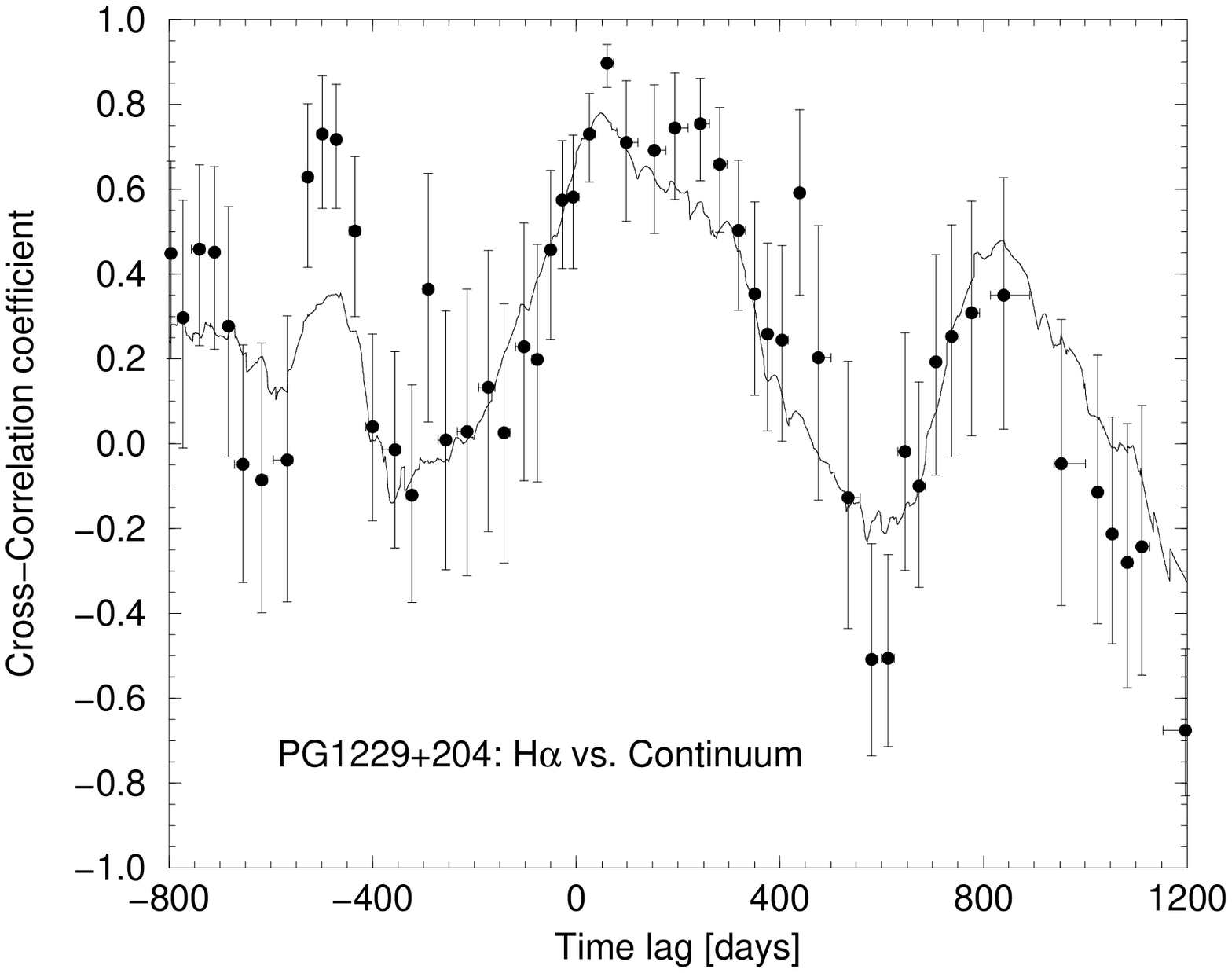}}
\vspace{-0.1cm}
\caption{ICCF {\it (solid line)} and ZDCF {\it (circles with error
bars)} for two cases. The two cross-correlation methods are in
excellent agreement. {\em Top panel:} PG\,0804+761 H$\beta$ cross
correlated with the optical continuum; the ICCF yields time lag of
$\Delta{t}({\rm centroid})=151^{+26}_{-24}$. {\em Bottom panel:}
PG\,1229+204 H$\alpha$ cross correlated with the optical continuum; the
ICCF time lag is $\Delta{t}({\rm centroid})=71^{+39}_{-46}$.}
\label{ccfs}
\end{figure}

In order to determine the time lags of the line light curves relative
to the continuum light curves we use two cross correlation methods:
one is the interpolated cross-correlation function (ICCF; Gaskell \&
Sparke 1986; Gaskell \& Peterson 1987; White \& Peterson 1994) where
one light curve is interpolated and then it is cross-correlated with
the second observed light curve; then the second light curve is
interpolate and cross-correlated with the first observed light curve.
The final ICCF is the mean of these two cross-correlation functions.
The second method is the $z$-transformed discrete correlation function
(ZDCF) of Alexander (1997) which is an improvement on the discrete
correlation function (DCF) of Edelson \& Krolik (1988). This method
applies Fisher's $z$ transformation to the correlation coefficients,
and uses equal population bins rather than the equal time bins used in
the DCF. The two independent methods are in excellent agreement for our
data and in the following analyses only the ICCF results are used.
Figure~\ref{ccfs} demonstrate the cross correlations of two of the line
light curves presented in Figure~\ref{light_curves} with their
corresponding continuum. For the purpose of this work we used the
centroid of the ICCF (computed from all points within 80\% of the ICCF
peak value) to define the time lag (Gaskell 1994 and references
therein).

To determine the uncertainties in the cross-correlation time lag we
used the model-independent FR/RSS Monte Carlo method of Peterson et al.
(1998b). In this method, each Monte Carlo simulation is composed of two
parts: The first is a ``random subset selection'' (RSS) procedure which
consists of randomly drawing, with replacement, from a light curve of
$N$ points a new sample of $N$ points. After the $N$ points are
selected, the redundant selections are removed from the sample such
that the temporal order of the remaining points is preserved. This
procedure accounts for the effect that individual data points have on
the cross-correlation. The second part is ``flux randomization'' (FR)
in which the observed fluxes are altered by random Gaussian deviates
scaled to the uncertainty ascribed to each point. This procedure
simulates the effect of measurement uncertainties. Applying the above
procedure to the line and continuum light curves and cross-correlating
them is considered one realization of the Monte Carlo simulations.
Using such $\sim 10000$ realizations builds up a cross-correlation peak
distribution (CCPD; Maoz \& Netzer 1989). The range of uncertainty
contains 68\% of the Monte Carlo realizations in the CCPD and thus
would correspond to $1\sigma$ uncertainty for a normal distribution.
Peterson et al. (1998b) demonstrate that under a wide variety of
realistic conditions, the combined FR/RSS procedure yields conservative
uncertainties compared to the true situation.

From the 46 line light curves 40 result with a significant correlation
(peak correlation coefficients above 0.4). All 40 CCFs indicate a
positive time lag of the Balmer lines with respect to the optical
continuum. The time lags are of order of a few weeks to a few months,
and the CCF peaks are highly significant for most lines. We conclude
that a time lag has been detected in one or more of the Balmer lines
for {\em all} 17 quasars.

\section{Size, Mass, and Luminosity -- Determination and Relations}

To construct the largest sample with available reverberation mapping
data we analyze our data together with comparable published data for
other AGNs. Wandel et al. (1999) have uniformly analyze reverberation
mapping data of 17 Seyfert 1's and deduce time lags using the same
techniques described in the previous section. Combining their results
with ours we obtain reliable size-mass-luminosity relations for 34 AGNs
spanning over 4 orders of magnitude in luminosity.

{\bf BLR Size:} 
Since we have both H$\alpha$ and H$\beta$ time lags for many objects we
average the two to get a better estimate for the Balmer lines time lag.
We do not include H$\gamma$ in the mean since its light curves are very
noisy due to the small S/N of the line, and the uncertainty of the
H$\gamma$ time lag is consistent with zero in several cases, hence,
counting it in the mean will add noise into our results. The BLR size
is then computed as the mean of the H$\alpha$ and H$\beta$ time lags
divided by a factor of $1+z$ to account for the cosmological time
dilution. For the Seyfert 1's we use the H$\beta$ time lags from Wandel
et al. (1999) and correct them by the $(1+z)^{-1}$ factor.

{\bf AGN Luminosity:}
A major limitation in the luminosity determination is the lack of
knowledge about the ionizing continuum and the spectral energy
distribution (SED) of the objects in question. Much of the ionizing
continuum is emitted in the unobservable far-UV and there are still
unsolved fundamental issues concerning the shape of the continuum
(e.g., Zheng et al. 1997; Laor et al. 1997). Even in one of the best
studied AGNs, NGC\,5548, the SED is poorly determined (Dumont,
Collin-Souffrin, \& Nazarova 1998). Another complication is the
contribution of the host galaxy to the luminosity of the nucleus. Since
resolving these complications is an issue for an in-depth study we took
the simplified approach (following Wandel et al. 1999) of using the
monochromatic luminosity, $\lambda L_{\lambda}$, at 5100~\AA\ (rest
wavelength) as our luminosity measure (assuming deceleration parameter
$q_0=0.5$, Hubble constant of $H_0=75$~km\,s$^{-1}$, and zero
cosmological constant). The uncertainty in this value is taken to be
only due to the variation of each object and is represented by the
root mean square (rms) of the continuum light curve.

{\bf Central Mass:}
Estimation of the AGN's central mass is carried out by assuming
gravitationally dominated motions of the BLR clouds: $M\approx
G^{-1}v^2r$ (e.g., Gaskell 1988; Wandel et al. 1999; Peterson \& Wandel
1999). In this relation the radius, $r$, is the BLR size computed above
and the velocity, $v$, is estimated from the rest-frame FWHM of the
emission line. Because the broad emission lines of AGNs are composed of
a narrow component superposed on a broader components, a unique FWHM
determination is not straightforward. We took two approaches to measure
the FWHM: the first approach is to measure the FWHM of the Balmer
lines in each spectrum for a given object and then to use the mean
FWHM, $v_{\rm FWHM}$(mean). The second approach is the one proposed by
Peterson et al. (1998a) of using the rms spectrum to compute the FWHM
of the lines, $v_{\rm FWHM}$(rms). In principle, constant features in the
mean spectrum (such as narrow forbidden emission lines, narrow
components of the permitted emission lines, galactic absorptions, and
constant continuum and broad-line features) are excluded in this
method. The FWHM from the rms spectrum measures only the part of the
line that varies and thus corresponds to the BLR size measured from the
reverberation mapping. In the following we uses the two approaches
together in order to compare them.

Following the approach of averaging the H$\alpha$ and H$\beta$ time
lags, we also average the FWHM of the H$\alpha$ and H$\beta$ lines. To
calculate the mass we also introduced a factor of $\sqrt{3}/2$, to
account for velocities in three dimensions and for using half of the
FWHM. The virial ``reverberation'' mass is then:
\begin{equation}
M=1.464\times 10^5\left(\frac{R_{\rm BLR}}{\rm lt\,days}\right)
\left(\frac{v_{\rm FWHM}}{\rm 10^3\,km\,s^{-1}}\right)^{2}M_{\odot} \ \ .
\end{equation} 

In Table~\ref{prop} we present the above computed properties for all
34 AGNs in our combined sample.

\begin{table}

\begin{centering}

\caption{Sizes, Luminosities, \& Masses \label{prop}}
\begin{tabular}{lcccc}
\tableline
\tableline
{Object} 		&
{$R_{\rm BLR}$} 	&
{$\lambda L_{\lambda}$(5100\AA )} &
{$M$(mean)} 		&
{$M$(rms)} 		\\
{} 		&
{(lt-days)} 	&
{$10^{44}$\,ergs\,s$^{-1}$} &
{$10^7M_{\odot}$}	&
{$10^7M_{\odot}$}	\\
\tableline
3C\,120      &             $ 42^{+ 27}_{- 20}$ & $ 0.73\pm  0.13$ & $ 2.3^{+ 1.5}_{- 1.1}$ & $ 3.0^{+ 1.9}_{- 1.4}$ \\
3C\,390.3    &       $ 22.9^{+  6.3}_{-  8.0}$ & $ 0.64\pm  0.11$ & $34^{+11}_{-13}$ & $37^{+12}_{-14}$ \\
Akn\,120     &       $ 37.4^{+  5.1}_{-  6.3}$ & $ 1.39\pm  0.26$ & $18.4^{+ 3.9}_{- 4.3}$ & $18.7^{+ 4.0}_{- 4.4}$ \\
Fairall\,9         &       $ 16.3^{+  3.3}_{-  7.6}$ & $ 1.37\pm  0.15$ & $ 8.0^{+ 2.4}_{- 4.1}$ & $ 8.3^{+ 2.5}_{- 4.3}$ \\
IC\,4329A    &     $  1.4^{+  3.3}_{-  2.9}$ & $ 0.164\pm  0.021$ & $ 0.5^{+ 1.3}_{- 1.1}$ & $ 0.7^{+ 1.8}_{- 1.6}$ \\
Mrk\,79      &     $ 17.7^{+  4.8}_{-  8.4}$ & $ 0.423\pm  0.056$ & $ 5.2^{+ 2.0}_{- 2.8}$ & $10.2^{+ 3.9}_{- 5.6}$ \\
Mrk\,110     &       $ 18.8^{+  6.3}_{-  6.6}$ & $ 0.38\pm  0.13$ & $ 0.56^{+ 0.20}_{- 0.21}$ & $ 0.77^{+ 0.28}_{- 0.29}$ \\
Mrk\,335     &     $ 16.4^{+  5.1}_{-  3.2}$ & $ 0.622\pm  0.057$ & $ 0.63^{+ 0.23}_{- 0.17}$ & $ 0.38^{+ 0.14}_{- 0.10}$ \\
Mrk\,509     &       $ 76.7^{+  6.3}_{-  6.0}$ & $ 1.47\pm  0.25$ & $ 5.78^{+ 0.68}_{- 0.66}$ & $ 9.2^{+ 1.1}_{- 1.1}$ \\
Mrk\,590     &     $ 20.0^{+  4.4}_{-  2.9}$ & $ 0.510\pm  0.096$ & $ 1.78^{+ 0.44}_{- 0.33}$ & $ 1.38^{+ 0.34}_{- 0.25}$ \\
Mrk\,817     &     $ 15.0^{+  4.2}_{-  3.4}$ & $ 0.526\pm  0.077$ & $ 4.4^{+ 1.3}_{- 1.1}$ & $ 3.54^{+ 1.03}_{- 0.86}$ \\
NGC\,3227    &   $ 10.9^{+  5.6}_{- 10.9}$ & $ 0.0202\pm  0.0011$ & $ 3.9^{+ 2.1}_{- 3.9}$ & $ 4.9^{+ 2.6}_{- 4.9}$ \\
NGC\,3783    &     $  4.5^{+  3.6}_{-  3.1}$ & $ 0.177\pm  0.015$ & $ 0.94^{+ 0.92}_{- 0.84}$ & $ 1.10^{+ 1.07}_{- 0.98}$ \\
NGC\,4051    & $  6.5^{+  6.6}_{-  4.1}$ & $ 0.00525\pm  0.00030$ & $ 0.13^{+ 0.13}_{- 0.08}$ & $ 0.14^{+ 0.15}_{- 0.09}$ \\
NGC\,4151    &   $  3.0^{+  1.8}_{-  1.4}$ & $ 0.0720\pm  0.0042$ & $ 1.53^{+ 1.06}_{- 0.89}$ & $ 1.20^{+ 0.83}_{- 0.70}$ \\
NGC\,5548    &     $ 21.2^{+  2.4}_{-  0.7}$ & $ 0.270\pm  0.053$ & $12.3^{+ 2.3}_{- 1.8}$ & $ 9.4^{+ 1.7}_{- 1.4}$ \\
NGC\,7469    &     $  4.9^{+  0.6}_{-  1.1}$ & $ 0.553\pm  0.016$ & $ 0.65^{+ 0.64}_{- 0.65}$ & $ 0.75^{+ 0.74}_{- 0.75}$ \\
PG\,0026+129 &               $113^{+ 18}_{- 21}$ & $ 7.0\pm  1.0$ & $ 5.4^{+ 1.0}_{- 1.1}$ & $ 2.66^{+ 0.49}_{- 0.55}$ \\
PG\,0052+251 &               $134^{+ 31}_{- 23}$ & $ 6.5\pm  1.1$ & $22.0^{+ 6.3}_{- 5.3}$ & $30.2^{+ 8.8}_{- 7.4}$ \\
PG\,0804+761 &               $156^{+ 15}_{- 13}$ & $ 6.6\pm  1.2$ & $18.9^{+ 1.9}_{- 1.7}$ & $16.3^{+ 1.6}_{- 1.5}$ \\
PG\,0844+349 &       $ 24.2^{+ 10.0}_{-  9.1}$ & $ 1.72\pm  0.17$ & $ 2.16^{+ 0.90}_{- 0.83}$ & $ 2.7^{+ 1.1}_{- 1.0}$ \\
PG\,0953+414 &               $151^{+ 22}_{- 27}$ & $11.9\pm  1.6$ & $18.4^{+ 2.8}_{- 3.4}$ & $16.4^{+ 2.5}_{- 3.0}$ \\
PG\,1211+143 &             $101^{+ 23}_{- 29}$ & $ 4.93\pm  0.80$ & $ 4.05^{+ 0.96}_{- 1.21}$ & $ 2.36^{+ 0.56}_{- 0.70}$ \\
PG\,1226+023 &               $387^{+ 58}_{- 50}$ & $64.4\pm  7.7$ & $55.0^{+ 8.9}_{- 7.9}$ & $23.5^{+ 3.7}_{- 3.3}$ \\
PG\,1229+204 &             $ 50^{+ 24}_{- 23}$ & $ 0.94\pm  0.10$ & $ 7.5^{+ 3.6}_{- 3.5}$ & $ 8.6^{+ 4.1}_{- 4.0}$ \\
PG\,1307+085 &             $124^{+ 45}_{- 80}$ & $ 5.27\pm  0.52$ & $28^{+11}_{-18}$ & $33^{+12}_{-22}$ \\
PG\,1351+640 &             $227^{+149}_{- 72}$ & $ 4.38\pm  0.43$ & $ 4.6^{+ 3.2}_{- 1.9}$ & $ 3.0^{+ 2.1}_{- 1.3}$ \\
PG\,1411+442 &             $102^{+ 38}_{- 37}$ & $ 3.25\pm  0.28$ & $ 8.0^{+ 3.0}_{- 2.9}$ & $ 8.8^{+ 3.3}_{- 3.2}$ \\
PG\,1426+015 &             $ 95^{+ 31}_{- 39}$ & $ 4.09\pm  0.63$ & $47^{+16}_{-20}$ & $37^{+13}_{-16}$ \\
PG\,1613+658 &             $ 39^{+ 20}_{- 14}$ & $ 6.96\pm  0.87$ & $24.1^{+12.5}_{- 8.9}$ & $ 2.37^{+ 1.23}_{- 0.88}$ \\
PG\,1617+175 &             $ 85^{+ 19}_{- 25}$ & $ 2.37\pm  0.41$ & $27.3^{+ 8.3}_{- 9.7}$ & $15.4^{+ 4.7}_{- 5.5}$ \\
PG\,1700+518 &               $ 88^{+190}_{-182}$ & $27.1\pm  1.9$ & $ 6^{+13}_{-13}$ & $ 5.0^{+11}_{-10}$ \\
PG\,1704+608 &               $319^{+184}_{-285}$ & $35.6\pm  5.2$ & $ 3.7^{+ 3.1}_{- 4.0}$ & $ 0.75^{+ 0.63}_{- 0.81}$ \\
PG\,2130+099 &             $200^{+ 67}_{- 18}$ & $ 2.16\pm  0.20$ & $14.4^{+ 5.1}_{- 1.7}$ & $20.2^{+ 7.1}_{- 2.4}$ \\
\tableline
\end{tabular}

\end{centering}

\end{table}

\subsection{Size--Luminosity Relation}

\begin{figure}
{\epsfxsize=13cm\epsfbox{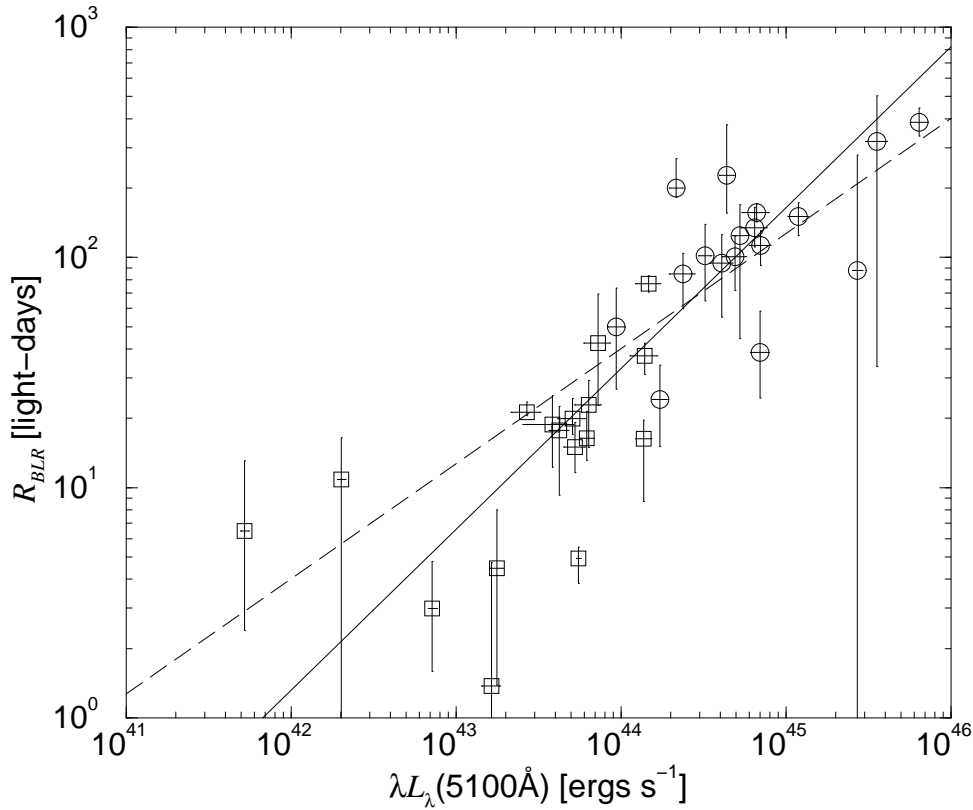}}
\caption{BLR size--luminosity relation. PG quasars are denoted by
circles and Seyfert 1's are denoted by squares. The solid line is the
best fit to the data. The dashed line is a fit with a slope of 0.5\,.}
\label{rvsl}
\end{figure}

The BLR size versus the luminosity is plotted in Figure~\ref{rvsl}.
The correlation coefficient is 0.827, and its significance level is
1.7$\times10^{-9}$. A linear fit to the points gives
\begin{equation}
R_{\rm BLR} = \left(32.9^{+2.0}_{-1.9}\right)\left(\frac{\lambda
L_{\lambda}(5100{\rm\AA} )}{\rm 10^{44}\,ergs\,s^{-1}}
\right)^{0.700\pm 0.033}\,{\rm lt\,days}
\end{equation}
(solid line plotted in Figure~\ref{rvsl}). Considering the Seyfert
nuclei ($\log(\lambda L_{\lambda}$(5100 \AA ))$\la 44.2$) or the PG
quasars alone, we find only marginally significant correlations,
probably because of the narrow luminosity ranges. A significant
correlation emerges only when using the whole luminosity range.

The relation we find between the BLR size and the luminosity does not agree
with earlier studies which found a smaller power-law index (closer to
0.5, e.g., Koratkar \& Gaskell 1991b; Wandel et al. 1999). A line with
0.5 slope was fitted to the data and is shown as a dashed line in
Figure~\ref{rvsl}. The combined sample is clearly inconsistent with
this slope.

Also, under the assumptions that the shape of the ionizing continuum in
AGN is independent of the luminosity, and that all AGNs are
characterized by the same ionization parameter and BLR density (as
suggested by the similar line ratios in low- and high-luminosity
sources), one expects $R_{\rm BLR}\propto L^{0.5}$. This theoretical
prediction is based on the assumption that the gas distribution, and
hence the mean BLR size, scales with the strength of the radiation
field. Our present result suggests that those assumptions should be
re-examined. In particular if we keep the assumption that the BLR
density is the same for all AGNs then the ionization parameter, $U$,
should have the relation $U\propto L^{-0.4}$.

\newpage

\subsection{Mass--Luminosity Relation}

\begin{figure}
{\epsfxsize=12.5cm\epsfbox{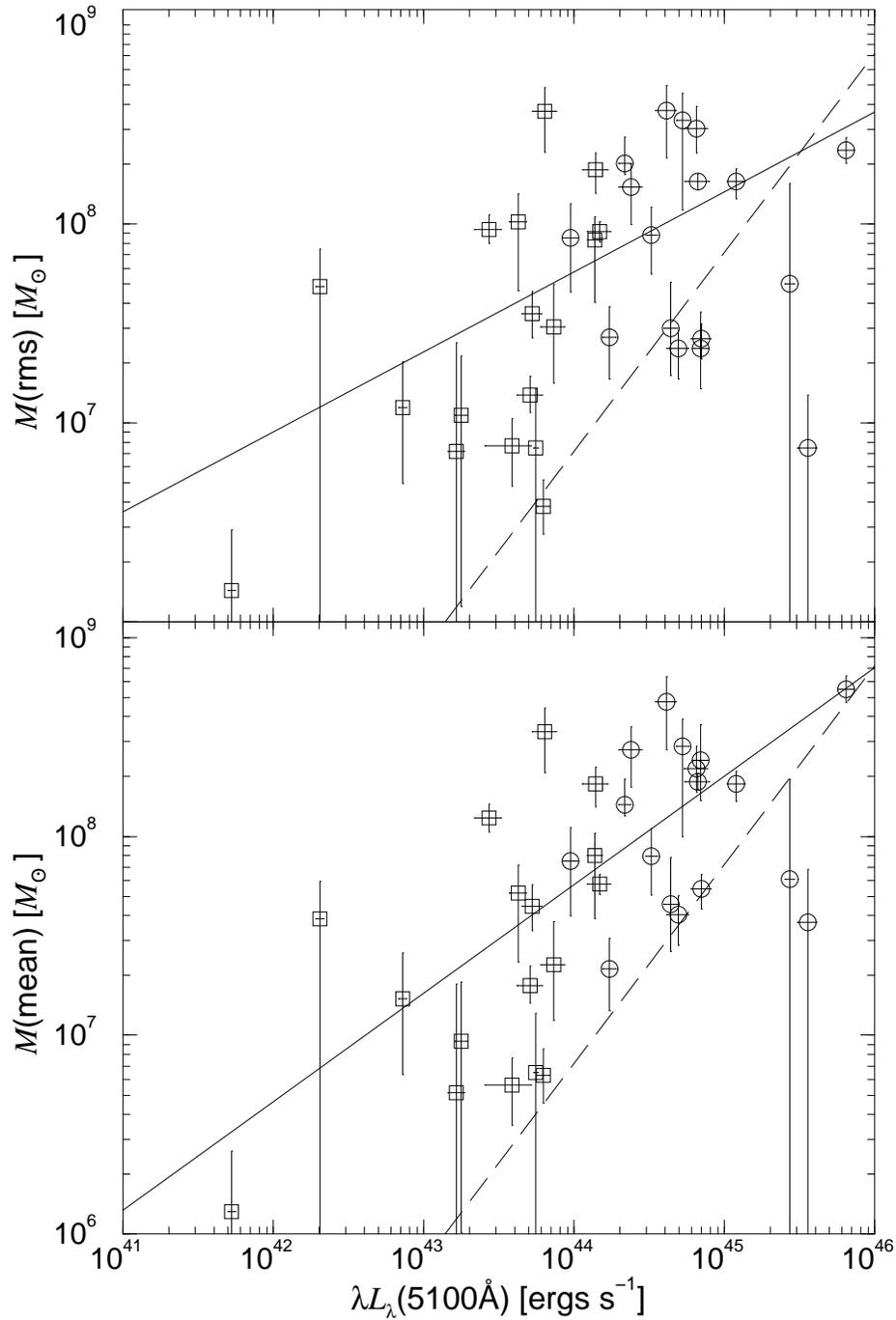}}
\caption{Mass--luminosity relations. Top: masses derived from
$v_{\rm FWHM}$(rms). Bottom: masses derived from $v_{\rm FWHM}$(mean).
Symbols as in Figure~\ref{rvsl}. Solid lines are the best fit to the
data. Dashed lines are the Eddington limit based on a rough estimate
for the bolometric luminosity (see text).}
\label{mvsl}
\end{figure}

The mass--luminosity relation is plotted in Figure~\ref{mvsl} for the
two mass estimates described above. Our mass estimates
based on the determination of the FWHM measured from the rms spectra
are plotted in the top panel. The correlation coefficient is 0.473
with a significance level of 4.7$\times10^{-3}$. A linear fit to this
relation gives
\begin{equation}
M({\rm rms}) = \left(5.75^{+0.39}_{-0.36}\right)\times10^7
\left(\frac{\lambda L_{\lambda}(5100{\rm\AA} 
)}{\rm 10^{44}\,ergs\,s^{-1}}\right)^{0.402\pm0.034}M_{\odot}
\end{equation}
and is plotted as a solid line in the diagram.

The mass estimates based on the determination of the FWHM from the mean
spectra are plotted versus luminosity in the bottom panel of
Figure~\ref{mvsl}. The correlation coefficient between these two
parameters is 0.646 and has a significance level of 3.7$\times10^{-5}$.
A linear fit gives
\begin{equation}
M({\rm mean}) = \left(5.71^{+0.46}_{-0.37}\right)\times10^7
\left(\frac{\lambda L_{\lambda}(5100{\rm \AA} 
)}{\rm 10^{44}\,ergs\,s^{-1}}\right)^{0.545\pm0.036}M_{\odot}
\end{equation}
and is also plotted as a solid line.

A surprising result is that when using the FWHM from the rms spectra
the mass-luminosity relation is less significant than when using the
mean FWHM. This is in contradiction to the theoretical considerations
leading to the use of the FWHM from the rms spectra (see above). This
disagreement can be attributed perhaps to the fact that the line fluxes
in the rms spectra are weaker and hence the uncertainty in the
corresponding FWHM might be larger.

Weighting the two mass--luminosity relations according to their
significance our results imply $M\propto L^{0.5\pm 0.1}$. This does not
agree with previous results -- for example Koratkar \& Gaskell (1991b)
found $M\propto L^{0.91\pm 0.25}$ and Wandel et al. (1999) reported on
$M\propto L^{0.77\pm 0.07}$. The fact that the scatter in the
mass--luminosity relation is larger than that of the size--luminosity
relation may indicate that luminosity, rather than mass, is the
variable that mainly determines the BLR size.

Using a rough estimate for the bolometric luminosity as $L_{\rm bol}$$\approx$\,$9\lambda L_{\lambda}$(5100\AA ), we obtain an Eddington ratio of
\begin{equation}
\frac{L_{\rm bol}}{L_{\rm Edd}}\approx 0.13\left(\frac{\lambda
L_{\lambda}(5100{\rm\AA} )}
{\rm 10^{44}\,ergs\,s^{-1}}\right)^{0.5}  \ \ .
\label{eqEddR}
\end{equation}
The Eddington limit, based on this rough estimate for $L_{\rm bol}$, is
plotted as a dashed line in Figure~\ref{mvsl}. This result indicates, for
the first time from reverberation mapping studies, that the Eddington
ratio increases with luminosity. Our findings are inconsistent with
theoretical models proposing the AGNs' luminosity to be a set fraction of
the Eddington luminosity, e.g., geometrically thin, optically thick
accretion disk model which implies $M\propto L$ (Laor \& Netzer 1989).
Our result suggests that the mass accretion rate grows with luminosity
much faster than the central mass, which would mean very different disk
properties in low- and high-luminosity sources. 

\newpage

\section{Reverberation Mapping of High-Luminosity Quasars}

Though the PG quasars are two orders of magnitude more luminous than
Seyfert 1 galaxies they are considered to be low-luminosity objects
among the known quasars. We can enlarge the luminosity range of AGNs
with reverberation mapping data by two orders of magnitude if we study
high-luminosity, high-redshift quasars ($L>10^{46}$\,ergs\,s$^{-1}$).
In fact, nowadays knowledge regarding line variability properties of
high-luminosity quasars resembles what was known for the PG quasars and
low-luminosity quasars in general about a decade ago (see
section~\ref{introduction}). There has been great progress in
reverberation mapping of Seyfert 1 galaxies and low-luminosity quasars
{\em but} nothing is known about line variability in high-luminosity
quasars. Do high-luminosity quasars emission lines respond to continuum
changes, as seen in Seyfert 1 galaxies and low-luminosity quasars? What
is the relative amplitude of the response, if any? What is the lag of
the response, reflecting the light-travel time across the BLR? Do
high-luminosity quasars BLR sizes scale with AGNs luminosity, and lie
on a continues relation from the faintest Seyferts to the brightest
quasars?

The best way to answer these questions is by monitoring
spectrophotometrically a sample of high-luminosity quasars. However,
the difficulties in carrying out such a program are enormous. As those
objects have apparent magnitude much fainter than the PG quasars one
needs a much larger telescope to monitor them (such as 8\,m class
telescope). Also, if the continuum variations and the BLR size are
scaling with the luminosity, much longer monitoring periods are
required (of order of 10 years -- though some preliminary results can
be obtained after only few years of observations).

\begin{figure}
\hglue-0.5cm{\epsfxsize=14.9cm\epsfbox{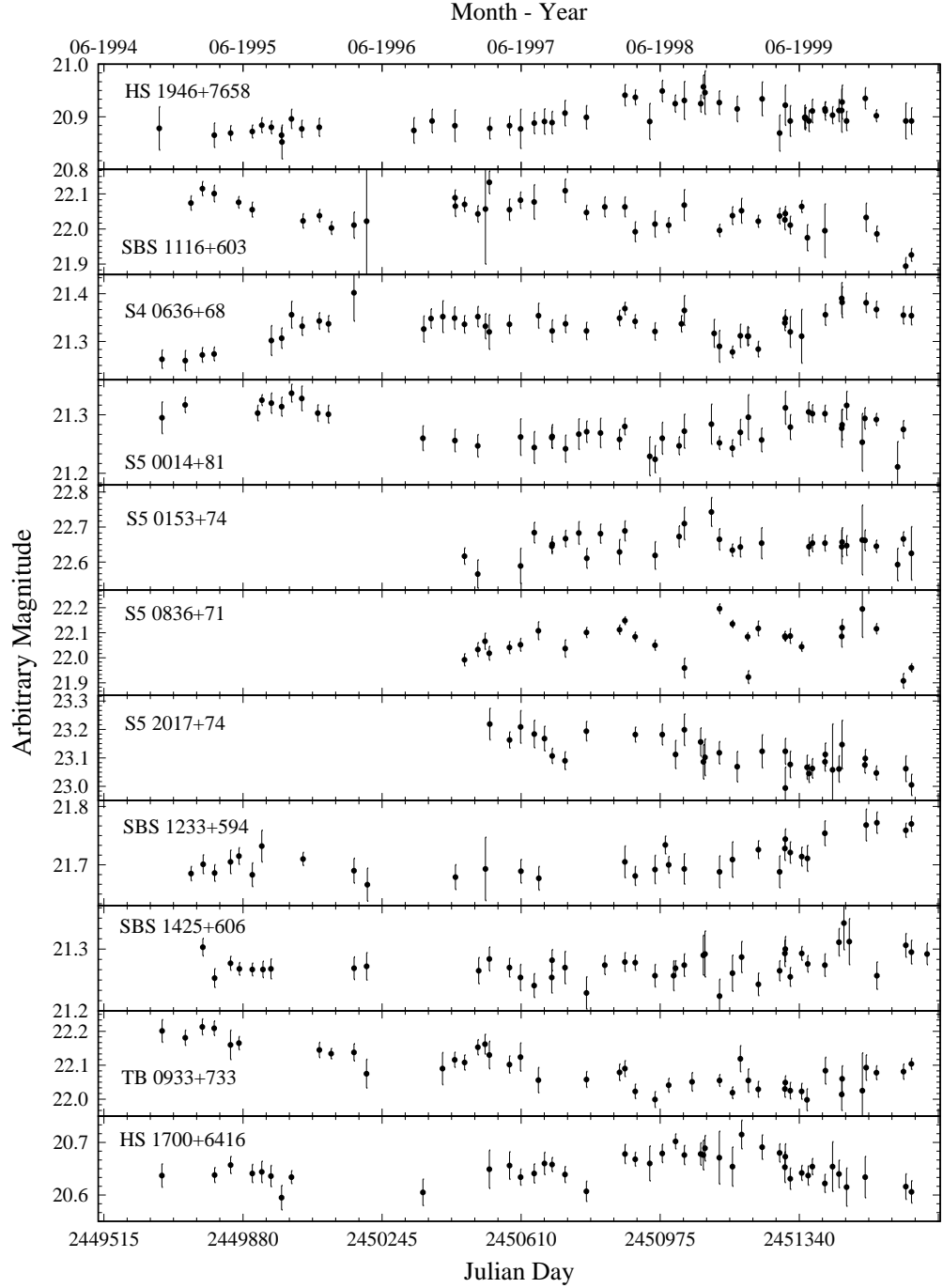}}
\vglue-0.5cm
\caption{R-band photometry light curves for 11 high-luminosity high-redshift quasars. All objects show variations of $\ga 0.1$ mag.}
\label{highl}
\end{figure}

In spite of the difficulties we initiated a program to monitor
high-luminosity quasars. Our sample consist of 11 quasars which were
chosen to have high northern declination, redshift of 2 to 3.4 (to
include the Ly$\alpha$ and C\,{\sc iv} UV lines in the optical region),
and observed $V$ magnitudes in the range of 16--18 mag. The sample is
being monitored photometrically in $B$ and $R$ bands each month at the
WO since 1994 November. Preliminary results are presented in
Figure~\ref{highl}. All 11 quasar show variations in the R-band flux of
$\ga 0.1$ mag. The quasar S5\,0836+71 has the largest variability which
is $\sim 0.3$ mag. The variation of the high-luminosity quasars are
smaller than the variations found for the PG quasars by about a factor
of 5 in magnitude. While in a comparable time-period the variations in
the PG quasars were in the range of 0.5--1 mag (Giveon et al. 1999) the
typical variations of the high-luminosity quasars are only 0.1 mag.
However, while the monitoring period is comparable in the observer time
frame the rest-frame monitoring period of the high-luminosity quasars
is about a factor of 4 smaller than the PG quasars monitoring period
and only amounts to $\sim$ 1.5 years.

\begin{figure}
\vspace{-0.2cm}
{\epsfxsize=12.7cm\epsfbox{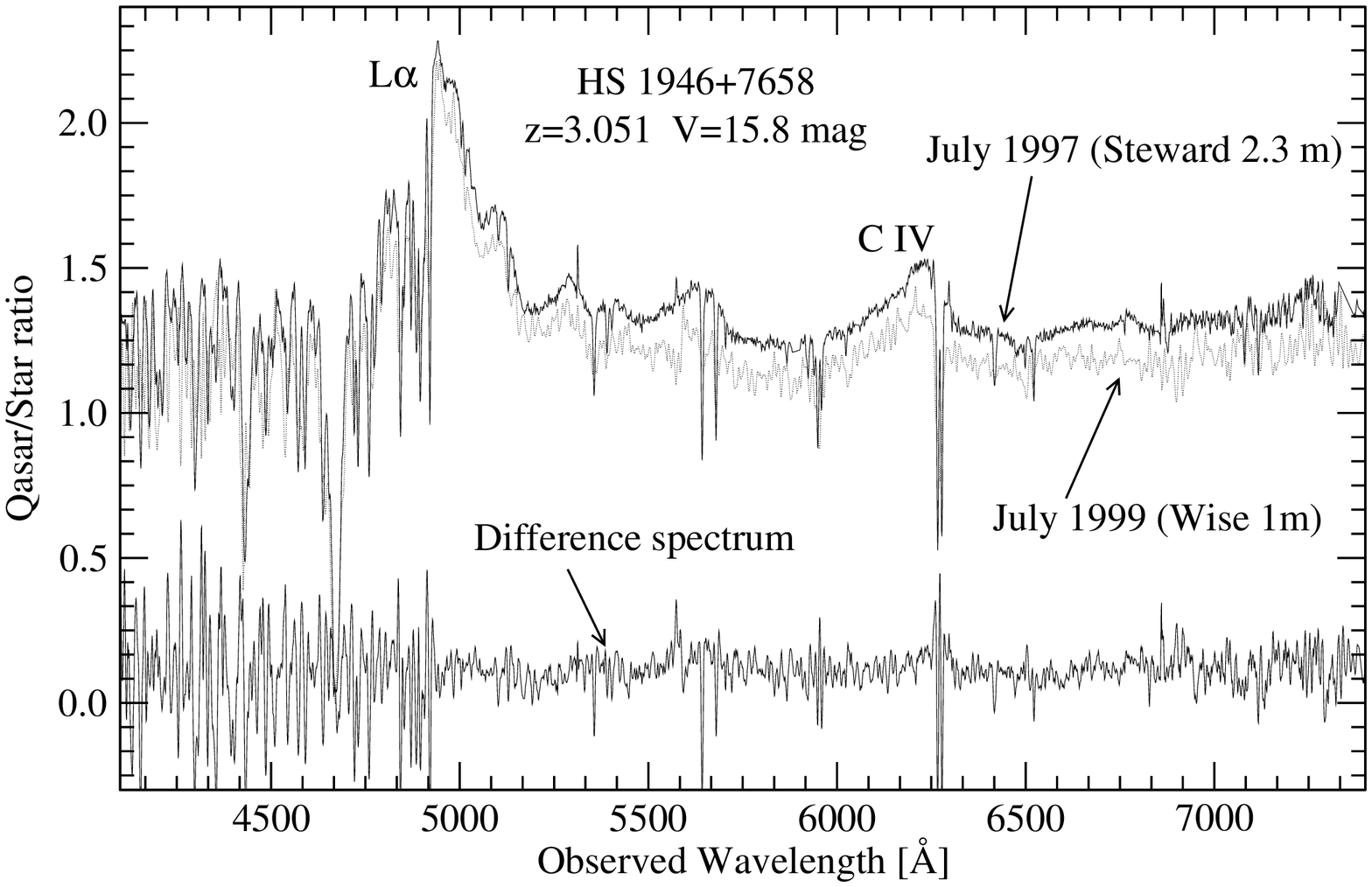}}
\vspace{-8.3cm}
\caption{Two spectra of HS\,1946+7658 ($z=3.051$, $V=15.8$~mag) taken
on 1997 July at SO 2.3\,m telescope (highest spectrum) and on 1999 July
at WO 1\,m telescope (dotted line spectrum). The difference spectrum is
shown at the bottom. Total integration time is 4 hours for each
spectrum. Though continuum variation is evident, no line variations were
found.}
\label{diff}
\vspace{-0.4cm}
\end{figure}

Spectrophotometric observations are needed in order to check a
corresponding variations in the emission lines. We carried out few
preliminary observations toward the brightest objects in the sample
using the WO and SO telescopes. Our best observations are demonstrated
in Figure~\ref{diff}. We present two observation epochs separated by
two years, each with total integration time of 4 hours. The observing
technique is, as described in section~\ref{observations}, of using a
comparison star in the slit simultaneously with the quasar. During the
2 years period a continuum variation of about 10\% is seen. This
corresponds to the fading this object shown in the R-band photometry
(Figure~\ref{highl}). No line variations is detected in the difference
of the two spectra.

Figure~\ref{diff}\,demonstrates\,that\,monitoring high-luminosity
high-redshift\,quasars is feasible and interesting results can be
obtained, however, it also demonstrates the difficulty to carry out such
a monitoring using small to medium size telescopes. Even though the
total integration time is 4 hours and the observing conditions are good
the S/N of the spectra is poor. It is clear that a large telescope is
needed to carry out this spectroscopic program. Early in 2000 we
started monitoring a subsample of our 11 quasars with the 8\,m
Hobby-Eberly telescope which is partly owned by the Pennsylvania State
University. We hope that within a short time we will be able to present
preliminary results from this long term monitoring program.

\section{Summary}

We reviewed the final results from a spectrophotometric monitoring of a
large, optically selected quasar sample (for the detailed study see
Kaspi et al. 2000). We find time lags between the optical continuum and
the Balmer-line light curves for {\em all} AGNs with adequate sampling.
Our work doubled the number of AGNs with a measured time lag (i.e., BLR
size). We also increased the available luminosity range for studying
the size--mass--luminosity relations in AGNs from two to four orders of
magnitude allowing, for the first time, to construct reliable
relations. Using all AGNs with known BLR size we find that the BLR
size scales with the 5100\,\AA\ luminosity as $L^{0.70\pm0.03}$. This
is significantly different from previous studies and is in
contradiction with simple theoretical expectations, both suggesting
$R_{\rm BLR}\propto L^{0.5}$. We also obtained a mass--luminosity relation
for AGNs, $M\propto L^{0.5\pm0.1}$\,, which, however, has a large
intrinsic scatter. These findings impose new and strict limitations on
theoretical considerations and constrain any theoretical models which
attempt to explain the AGN phenomenon.

One of our major current goals is to further increase our knowledge
about continuum and line variations and BLR size in high-luminosity
quasars ($L>10^{46}$\,ergs\,s$^{-1}$). To that end we are initiating a
program to monitor a sample of 11 high-luminosity high-redshift
quasars. We presented preliminary results of the continuum variations
in those objects to be $\ga 0.1$ mag. Spectrophotometric observations
of the sample is on the way and we hope that with 8\,m class telescopes
such as the Hobby-Eberly telescope we will be able in a few years to
extend our knowledge about BLR size to the most luminous quasars. The
conclusion of this long-term monitoring will hopefully complete the
reverberation studies to cover the entire AGNs luminosity range of
$10^{41}$--$10^{48}$\,ergs\,s$^{-1}$.

\acknowledgments
I deeply thank my collaborators in the PG quasars monitoring program
Uriel Giveon, Buell Jannuzi, Dan Maoz, Hagai Netzer, and Paul Smith who
spent years of work to achieve the results presented here. I would
also like to thank my additional collaborator in the high-luminosity
quasars monitoring program Niel Brandt, Donald Schneider, and Ohad
Shemmer. Research at the WO is supported by grants from the Israel
Science Foundation. Monitoring of PG quasars at SO was supported by
NASA grant NAG 5-1630.

\end{document}